\documentclass[%
 reprint,
superscriptaddress,
 amsmath,amssymb,
 aps,
]{revtex4-2}

\usepackage{graphicx}
\usepackage{dcolumn}
\usepackage{bm}
\usepackage[version=3]{mhchem}

\begin{document}

\preprint{APS/123-QED}

\title{Formation of lead halide perovskite precursors in solution: Insight from electronic-structure theory}

\author{Richard Schier}
\affiliation{Humboldt-Universit\"at zu Berlin, Physics Department and IRIS Adlershof, 12489 Berlin, Germany}
\author{Alejandro Conesa Rodriguez}
\affiliation{Humboldt-Universit\"at zu Berlin, Physics Department and IRIS Adlershof, 12489 Berlin, Germany}
\author{Ana M. Valencia}
\affiliation{Humboldt-Universit\"at zu Berlin, Physics Department and IRIS Adlershof, 12489 Berlin, Germany}
\affiliation{Carl von Ossietzky Universit\"at Oldenburg, Institute of Physics, 26129 Oldenburg, Germany}
\author{Caterina Cocchi}
\affiliation{Humboldt-Universit\"at zu Berlin, Physics Department and IRIS Adlershof, 12489 Berlin, Germany}
\affiliation{Carl von Ossietzky Universit\"at Oldenburg, Institute of Physics, 26129 Oldenburg, Germany}
\email{caterina.cocchi@uni-oldenburg.de}

\date{\today}

\begin{abstract}
Understanding the formation of lead halide (LH) perovskite solution precursors is crucial to gain insight into the evolution of these materials to thin films for solar cells. Using density-functional theory in conjunction with the polarizable continuum model, we investigate 18 complexes with chemical formula \ce{PbX2M4}, where X = Cl, Br, I and M are common solvent molecules. Through the analysis of structural properties, binding energies, and charge distributions, we clarify the role of halogen species and solvent molecules in the formation of LH perovskite precursors. We find that interatomic distances are critically affected by the halogen species, while the energetic stability is driven by the solvent coordination to the backbones. Regardless of the solvent, lead iodide complexes are more strongly bound than the others. Based on the charge distribution analysis, we find that all solvent molecules bind covalently with the LH backbones and that Pb-I and Pb-Br bonds lose ionicity in solution. Our results contribute to clarify the physical properties of LH perovskite solution precursors and offer a valuable starting point for further investigations on their crystalline intermediates.
\end{abstract}

\maketitle


\section{Introduction}
Lead halide perovskites (LHPs) are among the most promising materials for the next generation of photovoltaic and opto-electronic applications~\cite{ha+14aom,jeon+15nat,xiao+17natph,huan+19ees,anda+20aem}.
One of their greatest advantages is their ability to be synthesized and processed in solution~\cite{jeon+14natm,desc+14jpcl,zhao+14jpcl,nie+15sci,cho+17am}, thereby substantially reducing manufacturing costs for thin-film production~\cite{yi+19oe,ghos+20ami,zeng+20ees,cui+21aesr}.
The importance of solution chemistry in the synthesis and characterization of LHPs has stimulated dedicated research in the last few years~\cite{mcdo+18am,li+19small,jung+19csr}.
Many studies have contributed to clarify the role of solvent-solute interactions in determining the crystal structure and the properties of the grown films~\cite{fu+15jacs,petr+17jpcc,hami+17acsel,stev+17cm,fole+17jmca,fate+18cm,li+19cgd,dutt+19acsel,orto+19cgd,shar+20ma,more+20rsca}.
An intricate interplay between solubility, polarity, dielectric screening, and coordinating ability has been accounted responsible for the efficacy of different solvents to form complexes with LHPs~\cite{hami+17acsel,stev+17cm,tuta+20jpcc}.
Furthermore, the choice of the solvent turned out to be a critical parameter for the formation of intermediate phases~\cite{waka+14cl,li+16am,jo+16ami,cao+16jacs,petr+17jpcc,vasq+21jec} and even for the creation of defects in the final products~\cite{rahi+16chpch,wang+17ne}.

This vast body of experimental research has been supported by theoretical and computational work.
The ability of \textit{in silico} experiments to decouple the various effects in play has been exploited to interpret experimental data on the formation and the optical response of these systems~\cite{stev+17cm,radi+19acsaem,radi+20jpcl,radi+20jpcb,vale+21jpcl}.
Additionally, high-throughput screening has been proven as a reliable and cost-effective method to explore large configurational spaces in view of the synthesis and characterization of these complex materials~\cite{stev+17cm,gu+20joule}.

Motivated by these successful contributions from theory, in this work, we want to further explore the formation of LHP solution precursors adopting a physicists' perspective.
With state-of-the-art methods of \textit{ab initio} electronic-structure theory, we investigate the stability, the structural properties, and the charge distribution of charge-neutral LHP building blocks with chemical formula \ce{PbX2} (X = Cl, Br, I) bound to four molecules of six common solvents (M), namely \ce{C2H3N} (ACN), \ce{C3H7NO} (DMF), \ce{C2H6OS} (DMSO), \ce{C4H6O2} (GBL), \ce{C5H9NO} (NMP) and \ce{C4H6O3} (PC) -- see Figure~\ref{fig:solvents}.
Although it is known that in experimental samples different lead halide species coexist~\cite{radi+19acsaem,shar+20ma}, we choose to focus on the aforementioned systems, which give rise to remarkable fingerprints in optical absorption measurements~\cite{rahi+16chpch,shar+20ma,shar+20chpch}, to avoid accounting for additional electrostatic interactions with counterions to ensure charge neutrality.
Our goal is to disclose general trends related to the energetics, the structural properties, and the charge distribution of \ce{PbX2M4} solvated compounds that are useful to gain further understanding on the solution chemistry of LHPs and also to interpret the formation of intermediate phases in their evolution to crystalline thin films.
The approach adopted herein is complementary to computational-chemistry methods that are popular in the study of solution complexes, as well as to experiments in which these compounds are synthesized and characterized.

\section{Methodology}
\begin{figure}
    \centering
  \includegraphics[width=0.5\textwidth]{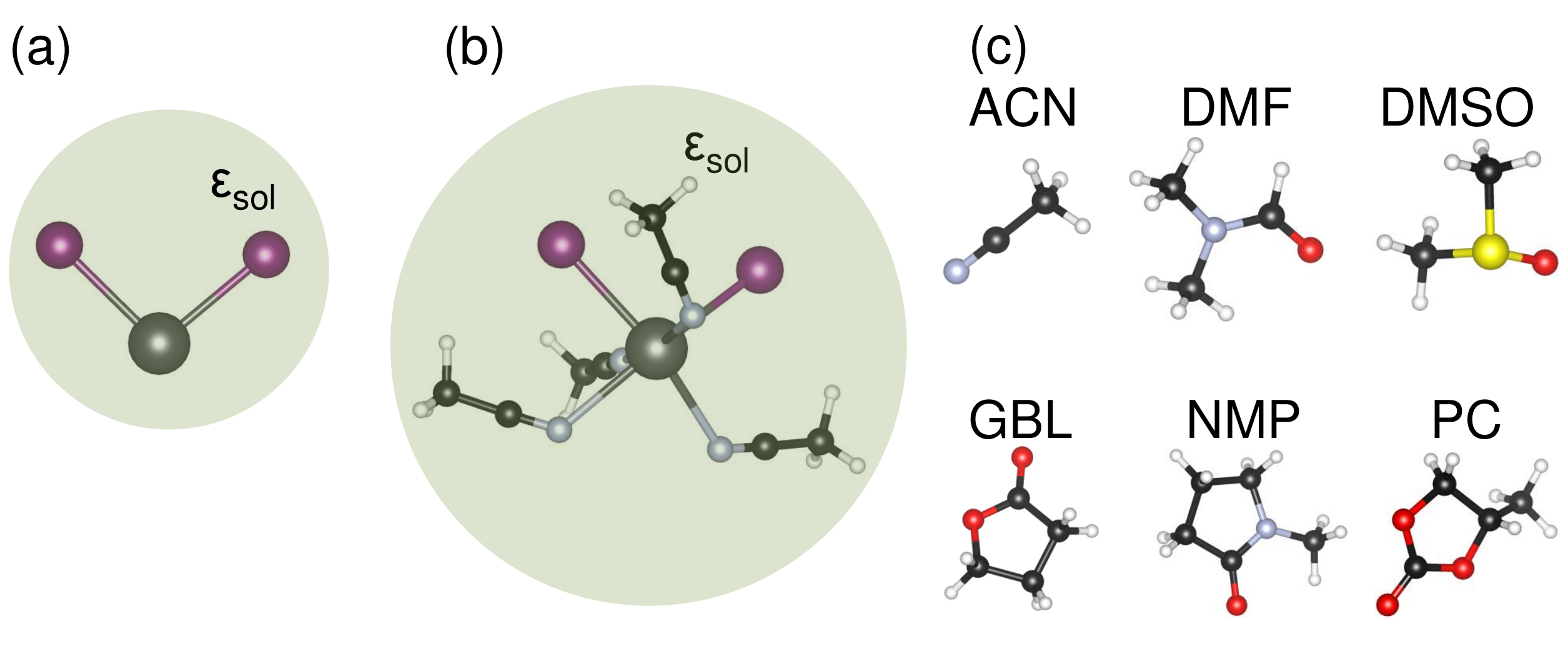}
	\caption{a) Sketch of \ce{PbX2} and b) \ce{PbX2ACN4} immersed in a solvent cavity; (c) Ball-and-stick representations of the solvent molecules considered in this work. Pb, I, O, S, C, H and N atoms are depicted in gray, purple, red, yellow, black, white and light blue, respectively.}
	\label{fig:solvents}
\end{figure}

The adopted \textit{ab initio} formalism based on density-functional theory (DFT)~\cite{hohe-kohn64pr,kohn-sham65pr,g16,grim+10jcp} with the range-separated hybrid functional CAM-B3LYP~\cite{yana+04cpl} (see computational details in the Supporting Information) enables an accurate modeling and an insightful analysis of physical mechanisms in play.
The representation of solute-solvent interactions through a hybrid atomistic/effective approach allows us to include both, the electronic interactions between \ce{PbX2} and the nearest solvent molecules, as well as the dielectric screening induced to the latter in the liquid environment.
In the considered structures, four solvent molecules are explicitly bound to the \ce{PbX2} compounds, thereby resulting in charge-neutral complexes with chemical formula \ce{PbX2M4} (M = ACN, DMF, DMSO, GBL, NMP, and PC).
These systems are embedded in a dielectric cavity described via the polarizable continuum model (PCM)~\cite{toma+05cr} coupled to DFT, which accounts for the screening induced by the surrounding solvent molecules (Figure~\ref{fig:solvents}b).
The following tabulated dielectric constants~\cite{wohlf15} are used to model the corresponding solvents: $\epsilon_{\text{DMSO}}$=46.7, $\epsilon_{\text{DMF}}$=38.2, $\epsilon_{\text{ACN}}$=37.5, $\epsilon_{\text{NMP}}$=32.5, $\epsilon_{\text{GBL}}$=41.7, $\epsilon_{\text{PC}}$=64.0.
Isolated \ce{PbX2} molecules in a solvent cavity with $\epsilon_{\text{ACN}}$=37.5 (see Figure~\ref{fig:solvents}a) are examined for comparison.
We cross-checked these calculations using $\epsilon_{\text{PC}}$=64.0 and we noticed no change in the considered quantities.

We construct all 18 systems assuming an axial geometry of PbX$_2$, which we found more stable in comparison with the equatorial X-Pb-X arrangement, in agreement with previous results on iodoplumbate compounds obtained at a similar level of theory~\cite{radi+19acsaem}.
By means of DFT calculations \textit{in vacuo}~\cite{blum+09cpc,havu+09jcp,perd+96prl,tkat-sche09prl,hirs77tca}, we checked that the enhanced stability of the compounds in the axial configuration is related to the more favorable charge distribution, which minimizes electron-electron repulsion between lead and halogen atoms (see Supporting Information, Figure~S1 and Table~S6, for further details).
Four solvent molecules are explicitly attached to each \ce{PbX2} backbone, thereby forming \ce{PbX2M4} complexes, which are optimized upon minimization of interatomic forces. 

\section{Results and Discussion}
\begin{figure*}
    \centering
  \includegraphics[width=\textwidth]{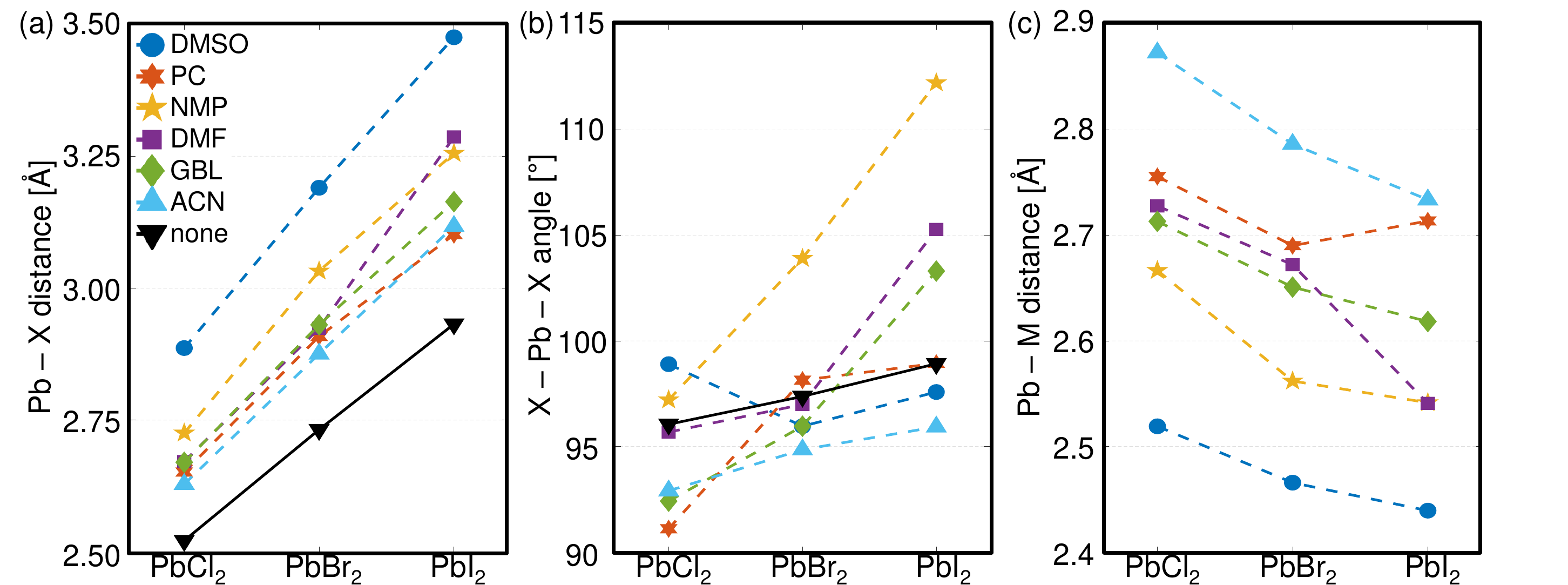}
	\caption{Structural properties including (a) the average distance between lead and halogen atoms, (b) the angle of the \ce{PbX2} backbone, and (c) the distance between lead atom and solvent  molecule (M). For comparison, results for \ce{PbX2} compounds without explicit solvent molecules (none) are reported in panels (a) and (b).}
	\label{fig:geometry}
\end{figure*}

The structural properties of the relaxed systems, summarized in  Figure~\ref{fig:geometry}, are consistent with the intuition that the distance between the Pb atom and the halogen species critically depends on the size of the latter: the smaller the halogen atoms, the shorter their separation from Pb.
The Pb-X distance is further enhanced by the presence of the explicit solvent molecules (Figure~\ref{fig:geometry}a).
On top of this, the dielectric screening of the implicit solvent cavity leads to a systematic and yet small increase of the Pb-X bond lengths on the order of 0.05~\AA{} (see Supporting Information, Table~S7), in agreement with previous findings on organic semiconductors in solution~\cite{krum+21pccp}.
Bonding with DMSO leads to the most sizeable increments in the Pb-X bond lengths, consistent with the relatively large steric hindrance of this molecule.
Likewise, the smallest changes with respect to the structure of \ce{PbX2} are induced by the smallest considered moiety, ACN.
In between, we find the trends induced by all the ringed compounds, such as GBL, NMP, and PC.
The behavior of \ce{PbX2DMF4} is noteworthy. 
In the complexes containing Cl and Br, Pb-X distances are almost identical to those with GBL.
On the other hand, Pb-I distances are enhanced by about 0.05~\AA{} by the presence of Pb-DMF bonds, as a result of the steric hindrance of this solvent molecule enhanced by the large atomic radius of iodine.
Our findings are in qualitative agreement with experimental reports of Pb-O distances in \ce{PbI2} solvated in DMF and DMSO~\cite{wu+14ees}.

For \ce{PbI2NMP4} and \ce{PbI2DMF4}, we can compare the computed bond lengths with available experimental references~\cite{shar+17cm,hami+20jpcc}.
Our values for Pb-I and Pb-O distances (both NMP and DMF are bound to \ce{PbI2} via O-bonds) agree well with the measurements~\cite{shar+17cm,hami+20jpcc}.
On average, our computed Pb-I bond lengths are consistent also with those measured for a crystalline phase of \ce{PbI2} in DMF~\cite{waka+14cl}, whereby in the inorganic backbone the lead atoms are connected with five iodine neighbors. 
In that structure, however, Pb-O distances are approximately 0.1~\AA{} lower compared to our estimates.
On the other hand, Pb-I bond lengths detected in an intermediate phase of methylammonium lead iodine in DMF~\cite{hao+14jacs} are only a few hundreds of \AA{} shorter than in our computational samples. 
This comparison suggests that the trends obtained in this study are relevant, at least qualitatively, also for the investigation of intermediate precursor phases of LHPs.

The variation of the X-Pb-X angle (see Figure~\ref{fig:geometry}b) is another indication of the non-trivial interplay between halogen species and solvent molecules in the final geometry of \ce{PbX2M4}. 
In the \ce{PbX2} systems simulated in the implicit solvent cavity (see Figure~\ref{fig:solvents}a), the X-Pb-X angle exhibits a mild increase with the size of the halogen atoms.
A similar trend but with systematically lower values is driven by the bonding of \ce{PbX2} with the smallest considered solvent molecule, ACN. 
On the other hand, bonding to NMP leads to a steep increase of the X-Pb-X angle in all systems, especially in \ce{PbBr2NMP4} and in \ce{PbI2NMP4}, where variations with respect to the ``bare'' \ce{PbBr2} and \ce{PbI2} counterparts exceed 6$^{\circ}$ and 13$^{\circ}$, respectively (see Supporting Information, Table~S2).
A monotonic enlargement of the X-Pb-X angle with increasing size of the halogen atom is also driven by PC, DMF, and GBL molecules.
Bonding with PC leads to a sizeable reduction of the Cl-Pb-Cl angle compared to the isolated \ce{PbCl2}, but to an insignificant variation in the compounds containing Br and I.
Conversely, DMF does not cause any noticeable variation in the Cl-Pb-Cl and Br-Pb-Br angles, while enhancing the I-Pb-I angle by more than 5$^{\circ}$ (see Figure~\ref{fig:geometry}b).
In the presence of GBL, the Cl-Pb-Cl angle is increased and the I-Pb-I angle is decreased, while the Br-Pb-Br one remains almost unaltered compared to \ce{PbBr2}.
Finally, bonding with DMSO leads to an increment of Cl-Pb-Cl angle compared to \ce{PbCl2} while to a reduction of the Br-Pb-Br and I-Pb-I angles.
In this case, the dielectric screening of the solvent cavity leads to lower values compared than \textit{in vacuo}, where the mutual interatomic repulsion is evidently larger (see Supporting Information, Tables~S6 and S7). 

Lastly, we analyze the average distance between the Pb atom and the solvent molecules in the considered complexes (see Figure~\ref{fig:geometry}c).
With the exceptions of PC, in all examined systems this length decreases with increasing size of the halogen species.
The solvent molecules are closest (furthest) from the central Pb atom in the presence of DMSO (ACN).
It is worth noting that ACN is the only solvent that is bound to Pb via a N atom; all other molecules are connected through a Pb-O bond. 

\begin{figure}
    \centering
  \includegraphics[width=0.4\textwidth]{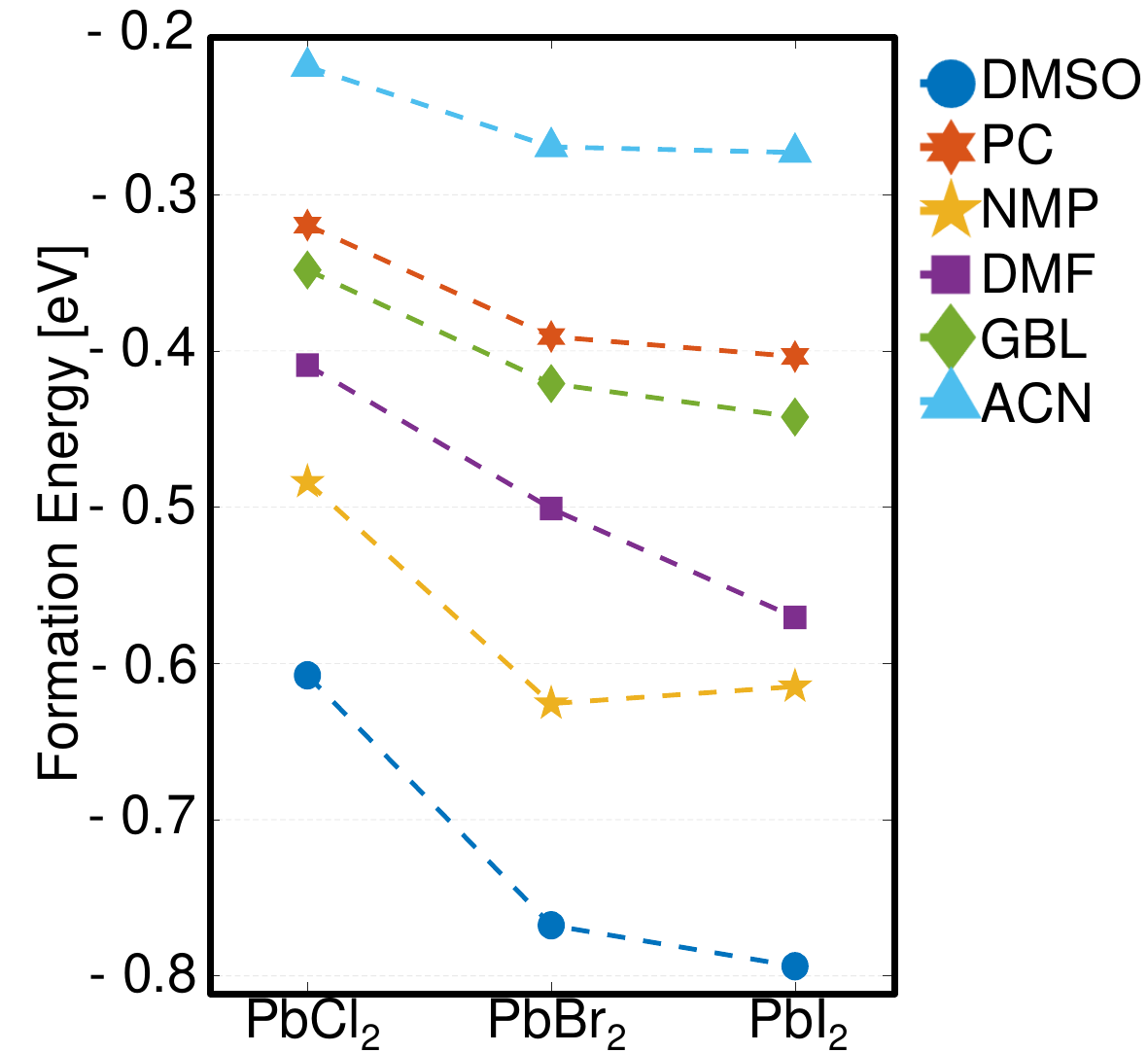}
	\caption{Formation energies of all the considered compounds.}
	\label{fig:formation}
\end{figure}

Next, we inspect the formation energies computed for all considered compounds in their optimized geometries analyzed above.
The formation energy, $E_f$, is evaluated as 
\begin{align}
 E_f = \dfrac{1}{4}(E_{tot} - E_{PbX_2} - E_{M_1} - E_{M_2} - E_{M_3} - E_{M_4}),
\label{eq:ef}
\end{align}
where $E_{tot}$ is the total energy of the relaxed \ce{PbX2M4} system, and $E_{PbX_2}$ and $E_{M_i}$ are the energies of the building blocks in their geometry within the complex.
With Eq.~\eqref{eq:ef}, we take into account separately the contribution of each individual solvent molecule in its own relaxed configuration, thereby accounting for variations among them when bonded to the \ce{PbX2} backbone.
Stable complexes give rise to negative values of $E_f$: the more negative this quantity, the more stable the compound.
Our results displayed in Figure~\ref{fig:formation} indicate that the most stable complexes are formed in the presence of DMSO, while the least stable ones include ACN.
The general trend indicates higher stability at increasing size of the halogen species with the only exception of the complexes in NMP, where the interaction with \ce{PbBr2} leads to a more negative formation energy.
It is worth noting that the description of the systems in an implicit solvent cavity leads to a substantial enhancement of the formation energy compared to \textit{in vacuo} conditions (see Table~S7 in the Supporting Information).

The results for the formation energies reported in Figure~\ref{fig:formation} are consistent with the coordination ability of the considered solvents in LHP solution precursors~\cite{li+15jacs,li+16am}, and are in line with previous theoretical predictions~\cite{radi+19acsaem,stev+17cm}.
A closer inspection of these data (see also Table~S3 in the Supporting Information) reveals, however, a few differences.
The values of formation energies reported in Ref.~\cite{radi+19acsaem} are systematically larger than those summarized in Figure~\ref{fig:formation}, with differences ranging from 0.05 to 0.10~eV that can be attributed to different computational parameters as well as to the slightly different formula adopted to estimate the formation energy.
In Ref.~\cite{stev+17cm}, the bonding with \ce{PbX2} is predicted to be more favorable with DMF rather than with NMP and, in general, values of formation energies are systematically larger than those calculated in the present work: this discrepancy can be again ascribed to the different approach for assessing the structural stability.

\begin{figure*}
    \centering
  \includegraphics[width=\textwidth]{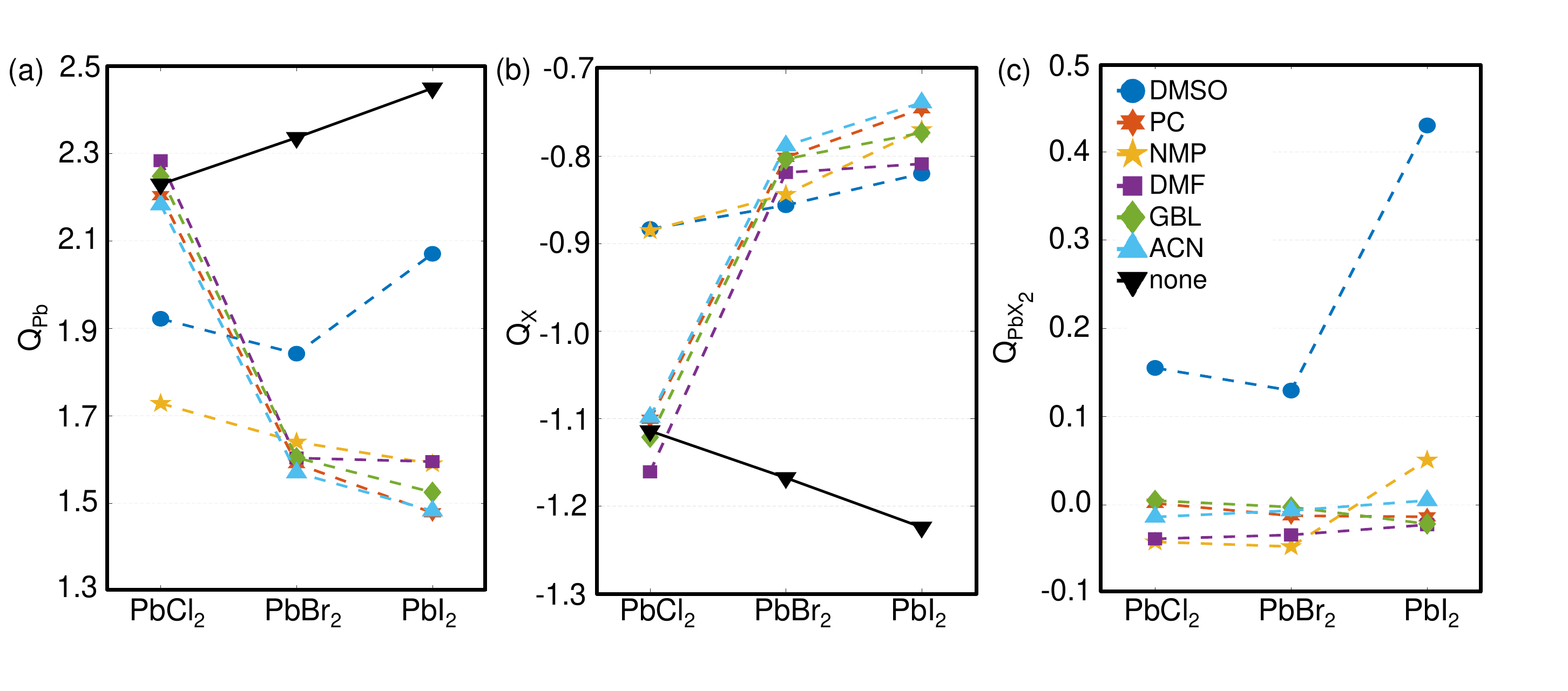}
	\caption{Bader partial charges (a) on the Pb atom, (b) averaged on the halogen atoms, and (c) on the \ce{PbX2} backbone. For comparison, results for \ce{PbX2} without explicit solvent molecules (none) are reported in panels (a) and (b).}
	\label{fig:charge}
\end{figure*}

We conclude our analysis by inspecting the charge distribution within the simulated complexes.
To do so, we make use of the Bader scheme~\cite{bade90oup,henk+06cms,sanv+07jcc,tang+09jpcm}, which enables a flexible partition of the charge density among different spatial domains. 
This method, based on the partition of the electron density computed from DFT is particularly useful in (hybrid) materials and interfaces formed by heterogeneous building blocks~\cite{niu+18,marc+19cm,vale+20pccp,jaco+20apx,schi+20jpcc}, and it is known to be more reliable than the Mulliken or the Hirshfeld schemes in systems with non-covalent bonds~\cite{saha+09ijqc}.
With the aim to understand the nature of the chemical bonds within the lead halide backbones as well as between them and the solvent molecules, we consider two types of partitions. 
First, we focus on the partial charges on Pb and on the halogen species in the \ce{PbX2M4} complexes (see Figure~\ref{fig:charge}a-b).
For comparison, \ce{PbX2} in an implicit solvent cavity without solvent molecules attached is also considered.
The results obtained for the latter systems confirm the net charge transfer between lead and halogen atoms and hence ionic nature of the Pb-X bond~\cite{borg+19ctc}.
The trend shown in Figure~\ref{fig:charge}b) indicates an increasingly negative charge in the less electronegative halogen atoms, with variations on the order of 0.1$e^-$. 
This qualitative finding is in agreement with earlier experimental observations~\cite{beng-holm89jcsfd}.

Examining now the results for the \ce{PbX2M4} complexes, we notice that the electronic interactions with the solvent molecules reduce the amount of negative charge on the heavier halogen species, thereby pointing to an enhancement of the covalent character of the Pb-I and Pb-Br bonds.
This effect is less pronounced with DMSO which is the considered solvent molecule with the largest number of atoms, whereby the resulting charge distribution is more faceted.
On the other hand, the character of the Pb-Cl bond is essentially unaltered by the interaction with the other solvent molecules, with the exceptions of DMSO and NMP.

The trends described above are reflected also in the total charge accumulated on the entire \ce{PbX2} unit, computed as the sums of the partial charges on lead and halogen atoms (see Figure~\ref{fig:charge}a-b) when interacting explicitly with the various solvent molecules (see Figure~\ref{fig:charge}c).
In most systems, there is an almost perfect cancellation between the partial charges on the solvent molecules and those on the lead halide backbone.
This cancellation is particularly efficient in presence of the electron-withdrawing solvents ACN, PC, and GBL.
Based on these results, we conclude that covalent bonds are formed between Pb and the solvent molecules in almost all considered systems.

The only apparent deviation from this trend occurs with DMSO.
Binding to this solvent molecule leaves the lead halide unit positively charged, with values between 0.1 and 0.2~$e^-$ in presence of Cl and Br and above 0.4~$e^-$ with iodine (see Table~S4 in the Supporting Information).
This result can be explained by considering the charge distribution occurring within DMSO. 
Already in the molecule, the oxygen lone pair generates a net charge close to -2~$e^-$ which is almost entirely compensated by the positive charge on the S atom.
When DMSO is bound to \ce{PbX2}, the charge on the O atom remains almost unaltered while the positive charge on S undergoes a decrease on the order of 0.2~$e^-$ (see Table~S5 in the Supporting Information).
As a result, each DMSO molecule in \ce{PbX2DMSO4} carries a negative charge contribution that sums up to -0.43~$e^-$, leaving the same amount of charge with opposite sign on the lead halide backbone (see Figure~\ref{fig:charge}c and Table~S4). 

\begin{figure*}
    \centering
  \includegraphics[width=\textwidth]{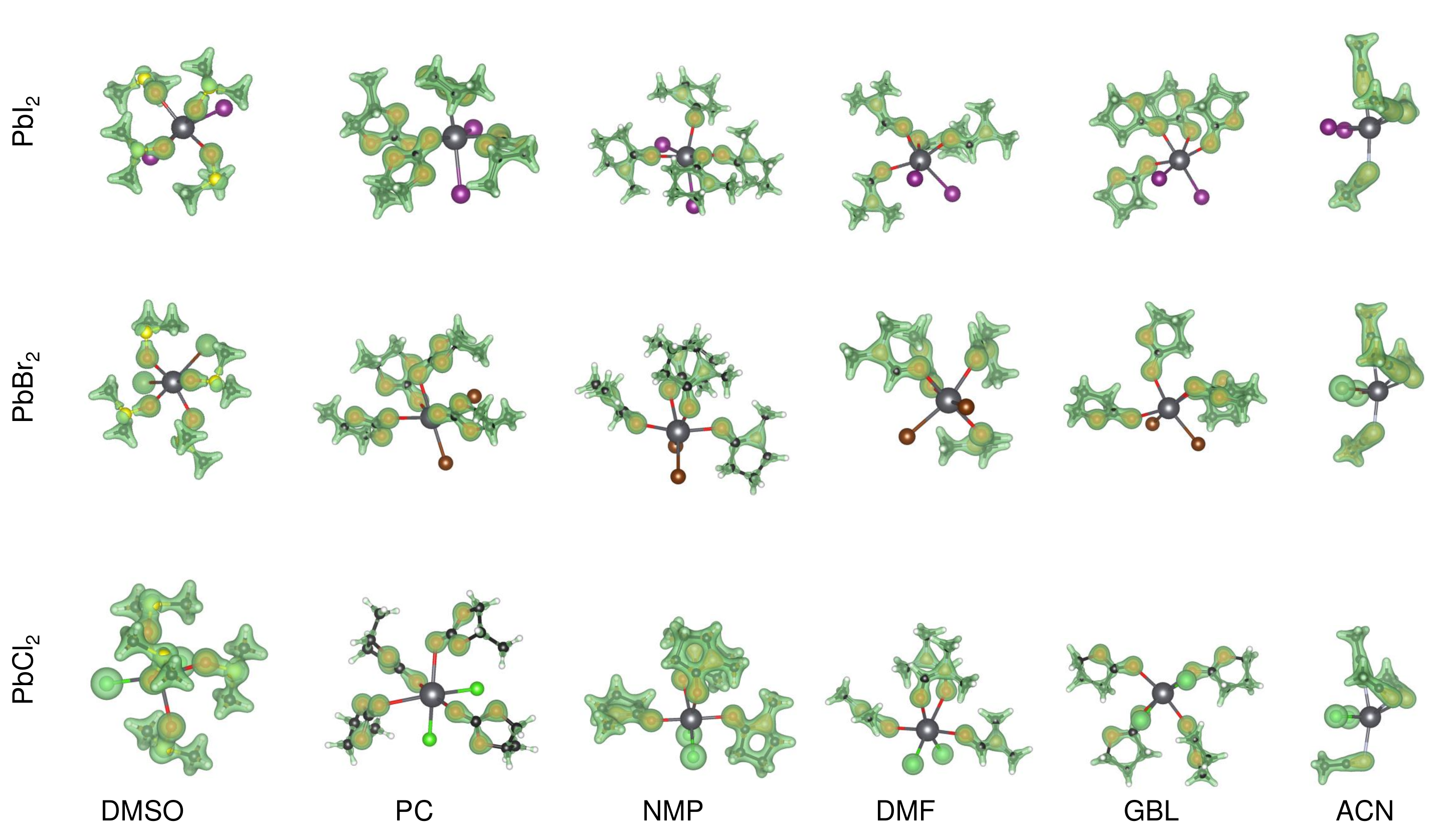}
	\caption{Total electronic charge distribution on all the considered compounds depicted with two isosurfaces corresponding to 4\% and 2\% of the maximum value. Pb, I, O, S, C, H and N atoms are depicted in gray, purple, red, yellow, black, white, and light blue, respectively. Graphics prepared with the VESTA software~\cite{vesta2011}.}
	\label{fig:isosurfaces}
\end{figure*}

We complement this analysis with the visual inspection of the total charge density on the \ce{PbX2M4} complexes, displayed in Figure\ref{fig:isosurfaces}.  
The largest contributions are visible on the solvent molecules and in particular on the ringed compounds DMF, GBL, and PC bound to \ce{PbCl2}, and on NMP interacting with \ce{PbBr2}.
Depending on the solvent and halogen species, we can distinguish between systems where the charge is mainly spread across the solvent and others where it is primarily localized on specific atoms or groups. 
A clear trend is visible depending on the size of the halogen atom: The charge becomes more (less) localized in the systems bound to PC (DMSO and NMP). 
In the presence of DMF, GBL and ACN, variations among compounds with different backbones are much smaller.
It is worth noting that in some cases, part of the charge is visible also on the halogen atoms, which is a signature of weaker charge transfer within the solvents. 
This is omnipresent in the considered compounds containing Cl, except when PC is involved.
At the isovalues plotted in Figure~\ref{fig:isosurfaces}, charge spills over onto Br only when \ce{PbBr2} is bound to DMSO and ACN.
Notably, no charge density appears on iodine atoms in \ce{PbI2}-based compounds.

\section{Summary and Conclusions}
In summary, we have presented a first-principles study of the formation of LHP solution precursors, focusing on charge-neutral systems with chemical formula \ce{PbX2M4} (X = Cl, Br, and I; M = DMSO, PC, NMP, DMF, GBL, and ACN).
The analysis of the structural properties of the obtained complexes reveals how the steric hindrance of the solvent molecules affects bond lengths and angles of the \ce{PbX2} backbone in the final geometries. 
Although the systems considered herein are only minimal units of actual LHP solution precursors, the general agreement of our computed bond lengths with those measured independently in various lead-iodide samples~\cite{waka+14cl,hao+14jacs,shar+17cm,hami+20jpcc}, even in more advanced stages of the evolution of LHPs to thin films, make it reasonable to anticipate that the disclosed effects are relevant also in subsequent stages of crystallization of these materials. 
The computed formation energies confirm the anticipated trends based on the coordination ability of the various solvents: DMSO leads to the most stable structures while ACN to the least favorable ones, regardless of the halogen in the backbone.
We also noticed a correlation between stability and size of the halogen species: Iodine-containing complexes are generally more stable than their counterparts with Cl or Br.
Our analysis of the charge distribution, based on the Bader partition scheme, offers important indications on the nature of the chemical bonds and on the corresponding influence of the solvent molecules. 
The ionic character of the Pb-I and Pb-Br bonds is considerably mitigated by the interaction with all the considered solvents. 
On the other hand, the ionicity of the Pb-Cl remains unaltered except in presence of the electron-donating solvents DMSO and NMP, which tend to reduce it.

Our study, performed from the perspective and with state-of-the-art methods of electronic-structure theory, complements the extended body of chemical studies on the solution chemistry of LHPs.
The presented quantum-mechanical analysis on the structural, energetic, and charge distribution properties of the considered compounds contributes to a deeper understanding of the fundamental properties of the solution precursors of these complex materials, shedding light on the nature and the strengths of their electronic stability and couplings.
Our results offer therefore a quantitative basis for future investigations on (pseudo)crystalline intermediate phases of LHPs in their evolution to thin-films for photovoltaic applications.

\section*{Supporting Information}
In the Supporting Information, we report the computational details, as well as additional information about the structural properties of the investigated compounds and the effects of the implicit solvents. The raw data plotted in Figure~\ref{fig:geometry}, Figure~\ref{fig:formation}, Figure~\ref{fig:charge}, and Figure~\ref{fig:isosurfaces} are also reported.

\section*{Acknowledgements}
We are grateful to Giovanni Procida, Holger-Dietrich Sa{\ss}nick, Oleksandra Shargaieva, and Eva Unger for fruitful discussions.
This work was supported by the German Research Foundation (DFG), Priority Programm SPP 2196 - Project number 424394788, by the German Federal Ministry of Education and Research (Professorinnenprogramm III), and from the State of Lower Saxony (Professorinnen für Niedersachsen).
A.C.R acknowledges financial support from the Erasmus+ Programme of the European Union.
Calculations were performed on the HPC cluster CARL at the
University of Oldenburg, funded by the DFG
(project number INST 184/157-1 FUGG) and by the Ministry
of Science and Culture of the Lower Saxony State.

\section*{Data Availability}
The data that support the findings of this study are openly available in Zenodo at 10.5281/zenodo.4767389.


\end{document}